\documentstyle[psfig,referee]{mn}

\title {PSR~J1740$-$3052 --- a pulsar with a massive companion}

\author[I. H. Stairs et al.]
{I. H. Stairs,$^{1,2}$
R. N. Manchester,$^3$ 
A. G. Lyne,$^1$
V. M. Kaspi,$^4$\thanks{On leave from Massachusetts Institute of 
Technology, Center for Space Research, 70 Vassar Street, Cambridge, 
MA~02139, USA}
F. Camilo,$^5$ 
\newauthor J. F. Bell,$^3$
N. D'Amico,$^{6,7}$ 
M. Kramer,$^1$
F. Crawford,$^8$\thanks{Present address: Management and Data Systems Division,
Lockheed Martin Corporation, P.O. Box 8048, Philadelphia, PA 19101, USA},
D. J. Morris,$^1$
\newauthor A. Possenti,$^{6}$ 
N. P. F. McKay,$^1$
S. L. Lumsden,$^9$
L. E. Tacconi-Garman,$^{10}$
\newauthor R. D. Cannon,$^{11}$
N. C. Hambly,$^{12}$
P. R. Wood$^{13}$ \\
$^1$University of Manchester,
Jodrell Bank Observatory, Macclesfield, Cheshire, SK11~9DL, UK \\
$^2$National Radio Astronomy
Observatory, P.O. Box 2, Green Bank, WV 24944, USA; email:
istairs@nrao.edu \\
$^3$Australia Telescope National Facility, CSIRO, P.O.~Box~76, 
Epping NSW~1710, Australia \\
$^4$Physics Department, 
McGill University, 3600 University Street, Montreal, Quebec, H3A~2T8, Canada \\
$^5$Columbia Astrophysics Laboratory, Columbia University,
550 W. 120th Street, New York, NY~10027, USA\\
$^6$Osservatorio Astronomico di Bologna, via Ranzani 1, 40127
Bologna, Italy \\
$^7$Instituto di Radioastronomia del CNR, via Gobetti 101, 40129
Bologna, Italy \\
$^8$ Massachusetts Institute of Technology, Center
for Space Research, 70 Vassar Street, Cambridge, MA~02139, USA \\
$^9$Department of Physics and Astronomy, E. C. Stoner Building, University of Leeds, Leeds, W. Yorkshire, L52~9JT, UK \\
$^{10}$ Max-Planck-Institut f\"ur Extraterrestrische Physik, Postfach 1312, 85741 Garching, Germany\\
$^{11}$Anglo-Australian Observatory, P.O.~Box 296, Epping NSW~1710, Australia \\
$^{12}$Institute for Astronomy, University of Edinburgh, Royal Observatory, 
Blackford Hill, Edinburgh, EH9 3HJ, UK \\
$^{13}$Research School of Astronomy \& Astrophysics, Institute of Advanced 
Studies, The Australian National University, \\
Cotter Road, Weston Creek, ACT~2611, Australia 
}

\begin{document}
\maketitle

\pagestyle{plain}
\begin{abstract}

We report on the discovery of a binary pulsar, PSR~J1740$-$3052,
during the Parkes multibeam survey. Timing observations of the 570-ms
pulsar at Jodrell Bank and Parkes show that it is young, with a
characteristic age of 350~kyr, and is in a 231-day, highly eccentric
orbit with a companion whose mass exceeds 11$\,M_{\odot}$. An accurate
position for the pulsar was obtained using the Australia Telescope
Compact Array.  Near-infrared 2.2-$\mu$m observations make with the
telescopes at the Siding Spring observatory reveal a late-type star
coincident with the pulsar position. However, we do not believe that
this star is the pulsar's companion, because a typical star of this
spectral type and required mass would extend beyond the pulsar's
orbit.  Furthermore, the measured advance of periastron of the pulsar
suggests a more compact companion, for example, a main-sequence star
with radius only a few times that of the sun.  Such a companion is
also more consistent with the small dispersion measure variations seen
near periastron.  Although we cannot conclusively rule out a
black-hole companion, we believe the companion is probably an early B
star, making the system similar to the binary PSR~J0045$-$7319.
\end{abstract}


\section{Introduction}\label{sec:intro}

Radio pulsars in binary systems provide a wealth of information about
neutron stars, their companions, and binary evolution.  Through radio
and optical observations of double-neutron-star and
neutron-star--white-dwarf binaries, an overall picture of binary
pulsar evolution has emerged (see, e.g., Bhattacharya and van den
Heuvel~1991 \nocite{bv91} and Phinney and Kulkarni~1994).
{\nocite{pk94} Young pulsars lose energy in the form of particles or
electromagnetic waves and are believed to spin down with approximately
constant magnetic field, eventually reaching the point at which the
radio emission mechanism is no longer effective.  An isolated neutron
star will never again be observed as a radio pulsar.  However, a
neutron star with a non-degenerate companion may be `recycled' by
accreting mass and orbital angular momentum from the evolving
companion in an X-ray binary phase.  The pulsar's spin period is
decreased, its magnetic field is reduced, and it begins once more to
emit as a radio pulsar.  There are several possible branches to this
evolution.  Low-mass companions permit stable mass transfer to the
neutron star over the lifetime of the giant phase, resulting in
pulsars with spin periods of a few milliseconds and low-mass He
white-dwarf companions.  Unstable mass transfer from higher-mass red
giants may yield pulsars with slightly longer spin periods and
heavier, CO white-dwarf companions.  A companion massive enough to
undergo a supernova explosion itself will result in either an unbound
pair of neutron stars, one somewhat recycled, one young, or a
double-neutron-star system such as PSR~B1913+16 \cite{ht75a}.

The precursors to these evolved binary systems must be young neutron
stars with non-degenerate companions, and some of these neutron stars
should be visible as radio pulsars.  Indeed, two such objects have
been reported to date: PSR~B1259$-$63, with a $\sim\!10\,M_{\odot}$
Be-star companion \cite{jml+92}, and PSR~J0045$-$7319, whose
companion is a B star also of mass $\sim\!10\,M_{\odot}$
\cite{kjb+94}.  Both of these main-sequence binaries are believed to
be progenitors of high-mass X-ray binaries (HMXBs).

In this paper, we report on the discovery of a third young radio
pulsar with a massive companion.  PSR~J1740$-$3052 is a 570\,ms
pulsar which is in a 231-day binary orbit with a companion of minimum
mass 11$\,M_{\odot}$.  This pulsar was discovered in the ongoing
Parkes multibeam pulsar survey, a large-scale survey for pulsars
currently being carried out using the 13-beam 1400-MHz receiver on the
Parkes 64-m radio telescope of the Australia Telescope National
Facility \cite{mlc+01}.  The discovery observations, radio
pulse timing and interferometric observations, and the results
obtained from them are described in \S\ref{sec:radio}.  Near-infrared
observations made to identify the pulsar companion are described in
\S\ref{sec:opt}. Radio and X-ray observations made around periastron
passages are described in \S\ref{sec:peri}. In \S\ref{sec:disc} we
discuss the implications of the observational results and the nature
of the pulsar companion.

\section{Radio Observations and Timing Solution}\label{sec:radio}

\subsection{Observations and Data Reduction}\label{sec:obs}

PSR~J1740$-$3052 was initially observed on 1997 Aug. 25, and the
survey reduction software identified it as a candidate with a 570-ms
period and dispersion measure (DM) of 739\,cm$^{-3}$\,pc.  Survey
parameters and procedures are described in detail by Manchester et
al. (2001). \nocite{mlc+01} The confirmation observation gave a period
which was substantially different from the initial discovery period,
indicating possible membership of a binary system.  The pulsar has
since been observed in a series of timing measurements at both the
Parkes 64-m telescope and the Lovell 76-m telescope at Jodrell Bank
Observatory, UK.

At Parkes, most data are recorded using the central beam of the
multibeam system, with a $2\times96\times3$ MHz filterbank centred on
1374\,MHz and a digital sampling rate of 250\,$\mu$s.  Details of the
timing observations may be found in Manchester et
al. (2001). \nocite{mlc+01} The pulsar is also frequently observed at
660 MHz, using a filterbank consisting of $2\times256\times0.125$ MHz
channels.

Observations at Jodrell Bank are made in a band centred near
1400\,MHz.  The filterbank consisted of $2\times32\times3$ MHz
channels centred on 1376\,MHz until 1999 Aug., and $2\times64\times1$
MHz channels centred on 1396\,MHz thereafter.

Data from both Parkes and Jodrell Bank are de-dispersed and folded at
the predicted topocentric pulsar period.  This process is performed
offline for the Parkes data and online for the Jodrell Bank data.
Total integration times per observation were typically 10\,min at
Parkes and 30\,min at Jodrell Bank.  Each pulse profile obtained by
summing over an observation is convolved with a high signal-to-noise
ratio ``standard profile'' producing a topocentric time-of-arrival
(TOA).  These are then processed using the {\sc TEMPO}
program\footnote{See http://pulsar.princeton.edu/tempo.}.  Barycentric
corrections are obtained using the Jet Propulsion Laboratory DE200
solar-system ephemeris \cite{sta82}.  The standard profiles at the two
observing frequencies are shown in Fig.~\ref{fig:profs}.

\subsection{Interferometric Position Determination}\label{sec:atca}

As PSR~J1740$-$3052 lies close to the ecliptic and is a member of a
long-period binary system, its position cannot yet be well determined
through standard pulsar timing analyses.  We therefore undertook
observations with the 6-element Australia Telescope Compact Array
(ATCA) on 1999 April 20.  Pulsar gating mode was used for simultaneous
observations at frequencies of 1384\,MHz and 2496\,MHz, with 128 MHz
bandwidth in both polarisations at each frequency.  The source
1934$-$638 was used to give the primary flux density calibration and
the three sources 1714$-$252, 1751$-$253 and 1830$-$360 were used as
phase calibrators.  The {\sc MIRIAD} software package\footnote{See
http://www.atnf.csiro.au/computing/software/miriad.} was used to
produce on- and off-pulse images of the pulsar field, and to fit a
point source to the differenced image. These images were made
separately using each of the phase calibrators at each of the two
frequencies, yielding five semi-independent determinations of the
position (the calibration source 1830$-$360 was resolved for long
baselines at 2496\,MHz and therefore was not usable at this frequency)
from which the position and its uncertainty were derived. This
position is listed in Table~\ref{tab:parms}. In these and other
entries in this table, the uncertainty is given in parentheses and
refers to the last quoted digit.

\subsection{Timing Results and Implications}\label{sec:soln}

Continued timing observations at Jodrell Bank and Parkes confirmed
that PSR~J1740$-$3052 is a member of a binary system in a highly
eccentric orbit.  Fig.~\ref{fig:perphase} shows the measured period
variation of PSR~J1740$-$3052 through the orbital period of 231 days.
Subsequently, we performed a phase-coherent analysis of the arrival
times over the period 1998 Aug. 7 -- 2000 Nov. 23, using the
main-sequence-star binary model of Wex (1998) in {\sc
TEMPO}. \nocite{wex98} The results are summarized in
Table~\ref{tab:parms}.  The errors are twice the formal $1\sigma$
values from the {\sc TEMPO} solution; we consider these to be
conservative estimates of the true $1\sigma$ errors.  Timing residuals
are shown in Fig.~\ref{fig:resids}.  The DM was obtained using the
1400\,MHz observations and five 660\,MHz observations at orbital
phases far from periastron.  Uncertainties in the pulse arrival times
were determined by requiring that the reduced $\chi^2$ be 1.0 for the
individual data sets from the different receivers and telescopes.
There are clearly some remaining systematic trends in the residuals.
As the orbital period is roughly 8 months and the sampling is
irregular, we are not yet able to fully distinguish between orbital
and annual effects in the data.  For this reason, we have fixed the
position at that determined by the ATCA observations, and find, in the
best simultaneous fit of all parameters, a marginally significant
result for the advance of periastron, $\dot \omega$.  If $\dot \omega$
is not included in the fit, the overall reduced $\chi^2$ increases
from 1.05 to 1.20; this increases our confidence in the significance
of the parameter.  With continued long-term timing, it should be
possible to refine the position as well as $\dot \omega$ and other
orbital parameters such as the change in projected semi-major axis,
$\dot x$, and the orbital period derivative, $\dot P_b$.

The spin parameters for PSR~J1740$-$3052 yield a characteristic age of
$\tau_c = P/2\dot P = 3.5\times10^5$\,yr, and a relatively high implied
surface dipole magnetic field strength of $B_0 = 3.2\times10^{19}(P \dot
P)^{1/2} = 3.9\times10^{12}$\,G.  The pulsar is clearly young and the
high magnetic field and long period suggest that it has not undergone
accretion of mass and angular momentum from its companion.

Using the Taylor \& Cordes~(1993)\nocite{tc93} model for the free
electron density distribution in the Galaxy and the measured DM, we
obtain an estimated distance for PSR~J1740$-$3052 of 11\,kpc, with a
nominal uncertainty of about 25\%.

A lower limit on the companion mass $m_2$ can be derived from the mass
function, assuming that the pulsar mass is $m_1 = 1.35\,M_{\odot}$, as
observed for neutron stars in binary radio pulsar systems \cite{tc99}:
\begin{equation}
f_1(m_1, m_2, i) = \frac{(m_2 \sin i)^3}{(m_1+m_2)^2} = 
 \frac{4 \pi^2}{T_{\odot}} \frac{x^3}{P_b^2} =
 \frac{4 \pi^2}{T_{\odot}} \frac{(a_1 \sin i)^3}{c^3\, P_b^2}, \label{eq:fmass}
\end{equation}
where $x=a_1 \sin i/c$ is the projected semi-major axis of the pulsar
orbit, $P_b$ is the orbital period, $c$ is the speed of light and
$T_\odot\equiv GM_\odot/c^3 = 4.925490947\,\mu$s.  Taking $\sin i$=1,
the minimum companion mass is $11\,M_{\odot}$, and the mass derived
using the median inclination angle of $i=60\degr$ is roughly
$16\,M_{\odot}$.  This suggests that the pulsar companion is either a
black hole or a non-degenerate star even more massive than the Be-star
companion (SS2883) of PSR~B1259$-$63.  If the companion is
non-degenerate, the high mass implies that it must be either an early
main sequence B~star or else a late-type supergiant.  We will consider
these two cases and that of a black hole in discussing the
implications of our observations, in the end coming to the conclusion
that a B~star is the most likely companion, although we have no direct
evidence for such a star at the pulsar's position.

An important point to consider is the size of the companion.  An early
B star of $11\,M_{\odot}$ is expected to have a radius of 6 or 7
$R_{\odot}$ (e.g., Cox 2000), \nocite{cox00} much smaller than the
pulsar orbit.  Of course, a black hole will also fit this criterion.
Late-type supergiants, on the other hand, have typical radii of
several hundred $R_{\odot}$ (e.g., van Belle et al. 1999),
\nocite{vlt+99} comparable to or larger than the pulsar's projected
orbital semi-major axis of 757 light-seconds, which is about 325
$R_{\odot}$ or 1.5\,AU.  The distance of closest approach of the two
bodies may be as small as 0.72\,AU, depending on the orbital
inclination angle.  However, to date, there has been no evidence of
eclipse of the pulsar at any phase of the orbit
(\S\ref{sec:peri_time}).  Furthermore, the
youth of the pulsar and eccentricity of the orbit indicate that no
significant mass has been transferred from the companion to the
pulsar.  Thus, if the companion is in fact a late-type supergiant, it
must be confined within its Roche lobe.  We use the formula of
Eggleton (1983) \nocite{egg83} to estimate the radius of the Roche
lobe of the companion star near periastron, arriving at roughly
0.4\,AU for stellar masses in the range $11\,M_{\odot}$ to
$16\,M_{\odot}$.  A late-type companion must therefore have a radius
smaller than this, unusually small for such a star.  On the assumption
that the star fills (or nearly fills) its Roche lobe, the lack of
eclipses requires the inclination angle to be $\la 70\degr$, resulting
in a companion mass of $\ga 12.5 M_{\odot}$.

\subsection{The Advance of Periastron}\label{sec:quad}

As mentioned in \S\ref{sec:soln}, the pulsar timing solution shows an
advance of periastron, $\dot \omega = 0.00021(7)^{\circ}\,$yr$^{-1}$.
This may be due to a combination of general-relativistic and classical
effects.  For a compact companion, general relativity predicts an
advance of
\begin{eqnarray}
\dot\omega & = & 3 \left(\frac{P_b}{2\pi}\right)^{-5/3} [T_\odot
  (m_1+m_2)]^{2/3}\,(1-e^2)^{-1} \nonumber \\
	& = &0.00023\;\degr {\rm yr}^{-1}, \label{eq:omdot_gr}
\end{eqnarray}
where $e$ is the orbital eccentricity and we assume a pulsar mass of
$m_1 = 1.35\,M_{\odot}$ and a companion mass of $m_2 = 16\,M_{\odot}$.
The observed value of $\dot\omega$ is comparable to this, so that a
black hole companion is certainly permitted by the observations.
Classical contributions to $\dot\omega$ and to $\dot x$, the
derivative of the projected orbital semi-major axis, may come from a
quadrupole moment of a companion star, induced either by the stellar
rotation or by tides raised by the neutron star
\cite{lbk95,kbm+96,wex98} .  For a stellar quadrupole moment $q$, the
expected classical contributions to $\dot\omega$ and $\dot x$ are
\cite{wex98}:
\begin{eqnarray}
\dot\omega & = & \frac{3 \pi q}{P_b a_1^2(1-e^2)^2}\left(1-\frac{3}{2}
	\sin^2\theta + \cot i\sin\theta\cos\theta\cos\Phi_0\right) \nonumber \\
 	   & = & 830\,\left(\frac{q}{{\rm AU}^2}\right)\,\sin^2i \left(1-\frac{3}{2}
	\sin^2\theta + \cot
	i\sin\theta\cos\theta\cos\Phi_0\right)\degr{\rm yr}^{-1}, 
        \label{eq:omdot} \\
\dot x & = & \frac{3 \pi q}{P_b c a_1^2(1-e^2)^2}(a_1 \sin i)\cot
i\sin\theta\cos\theta\sin\Phi_0 \nonumber \\
 	   & = & 3.5\times 10^{-4}\,\left(\frac{q}{{\rm AU}^2}\right)\, \sin^2 i \cot
	i\sin\theta\cos\theta\sin\Phi_0, \label{eq:xdot}
\end{eqnarray}
where $\Phi_0$ is the precession phase, and $\theta$ is the angle
between the orbital angular momentum and the star's spin vector or the
line perpendicular to the plane of the tidal bulge.

To estimate the relative importance of the spin and tidal quadrupoles,
$q_S$ and $q_T$, for a B main-sequence star, we follow the reasoning of Lai,
Bildsten \& Kaspi (1995) and take the ratio
\begin{equation}
\frac{q_S}{q_T} \sim \left(\frac{P_b}{P_S}\right)^2 \frac{(m_1+m_2)}{m_1}(1-e)^3, \label{eqn:quadrat}
\end{equation}
where $P_S$ is the spin period of the star, which we estimate to be
$2\times10^5$\,sec, similar to that for the B1V companion to
PSR~J0045$-$7319 \cite{bbs+95}.  From this, we derive $q_S/q_T \sim
10^4$; consequently the tidal quadrupole is negligible in
equations~\ref{eq:omdot}--\ref{eq:xdot} above.

The measured non-general-relativistic contribution to $\dot \omega$ is
$\la 0.00012\degr$\,yr$^{-1}$.  Assuming $i = 60\degr$, $\theta =
20\degr$ \cite{bai88} and $\Phi_0 = 45\degr$, we find $q_S \la 2.0
\times 10^{-7}$\,AU$^{2}$.  The rotationally-induced quadrupole
moment, $q_S$, is given by \cite{cow38}: 
\begin{equation}
q_S = \frac{2}{3}\frac{k R_2^{5} \Omega^2}{G m_2} \label{eq:qrot}
\end{equation}
where $G$ is Newton's constant, $\Omega=2\pi/P_S$ is the stellar
angular velocity, $m_2$ and $R_2$ are the companion mass and radius,
and $k$ is the apsidal constant representing the structure of the
star, estimated for such a star to be $\sim 0.01$ \cite{sch58,cg92b}.
With a radius estimate of 6.5\,$R_{\odot}$ as for the companion to
PSR~J0045$-$7319 \cite{bbs+95}, this leads to an estimate of the
stellar rotational angular velocity of $\Omega \la
2.7\times10^{-5}$\,rad\,s$^{-1}$, $P_S \ga 2.3\times10^5$\,s, consistent
with that used above in equation~\ref{eqn:quadrat}.

Based on this estimate of the spin quadrupole, the change in the
projected semi-major axis is predicted to be $\dot x \la 4.6\times
10^{-11}\, \sin^2 i \cot i\sin\theta\cos\theta\sin\Phi_0$, comparable
to our measured limit of $4 \times 10^{-11}$ and providing no
constraints on the inclination angle or precession phase.

In the case of a late-type companion, we estimate $R_2 \sim 0.35\,$AU
based on the Roche lobe argument in \S\ref{sec:soln}, $m_2 =
12.5\,M_{\odot}$, $i = 70\degr$ and $k \sim 0.03$ (A. Claret, private
communication).  The spin of the star, $\Omega$, is not well
known. Based on an estimate of $v\sin i
\simeq 1$\,km\,s$^{-1}$ for late supergiants \cite{dm99}, we arrive at
$\Omega \simeq 1.2\times10^{-7}$\,rad\,s$^{-1}$, and hence $q_S \simeq
3.0\times10^{-6}$\,AU$^{2}$ for a 12.5\,$M_{\odot}$ star. However, as
the radius of the star must be significantly smaller than that of most
supergiants, $v\sin i$ could well be larger.  An estimate of
$\Omega \simeq 3.1\times 10^{-7}$\,rad\,s$^{-1}$ and hence $q_S \simeq
2.0\times10^{-5}$\,AU$^{2}$ comes from setting the stellar rotational
rate equal to the orbital frequency.  

The quadrupole moment due to a tide, $q_T$, can be written as:
\begin{equation}
q_T = k \frac{m_1}{m_2} R_2^2 \left(\frac{R_2}{r}\right)^3
\label{eq:qstide}
\end{equation}
where $r$ is the centre-of-mass separation at any given time
\cite{lbk95}.  Under the same assumptions given above for a late-type 
companion, we find $q_T \simeq 5.3\times10^{-6}$\,AU$^{2}$ averaged
over the orbit.  Thus the static tidal and the rotational quadrupole
moments may well be roughly the same order of magnitude, although
probably not aligned.

The tidal quadrupole will not contribute to the value of $\dot x$ in
equation~\ref{eq:xdot} above, as the tidal bulge will be aligned with
the orbital plane.  The measured limit of $\dot x < 4.0\times
10^{-11}$, combined with the assumptions that $i = 70\degr$ and
$\theta = 20\degr$, yields a value of $q_S \la 1.6
\times 10^{-6}$\,AU$^{2}$ or $\Omega \la 8.0 \times 10^{-8}$\,rad\,s$^{-1}$ 
for $\Phi_0 = 45\degr$, slightly smaller than our estimate above.

However, both the static tidal and rotational quadrupole moments will
contribute to the value of $\dot \omega$, although, given the difference
in magnitudes and alignments, the contributions will differ in size
and perhaps also in sign.  Using the estimate of $q_S$ derived in the
preceding paragraph, we arrive at a maximum rotational quadrupole
contribution to $\dot \omega$ of $0.0013\;\degr {\rm yr}^{-1}$.  The
estimate of the static tidal contribution is larger, $\sim
0.0044\;\degr {\rm yr}^{-1}$.  The actual measured value of
$\dot\omega$ is $0.00021(7)\degr {\rm yr}^{-1}$, about an order of
magnitude smaller than either of these values.

With the simple assumptions we have made, it is difficult to reconcile
the observed and predicted values of $\dot \omega$ for a late-type
supergiant companion.  It may be possible to do so through fine-tuning
of the stellar spin, the precession phase and the spin inclination
angle, but such a solution seems unlikely.

\section{Near-IR Search for the Companion}\label{sec:opt}

In order to clarify the nature of the companion, we undertook
observations in other wavebands.  PSR~J1740$-$3052 lies only $0.\!\!\degr13$
from the Galactic plane, very close to the Galactic Centre and at a
distance of roughly 11\,kpc; large amounts of extinction are therefore
expected at optical wavelengths.  Indeed, there is no object at the
pulsar position in the Digitised Sky Survey (DSS), which covers the
$V$-band to approximately 16th magnitude.  There is also no sign of an
optical image on deeper Sky Survey photographs taken with the
U.K. Schmidt Telescope, to limiting magnitudes of 21 in $B$, 20 in $R$
and 18 in $I$.  We therefore carried out observations at infrared
wavelengths.

On 1999 May 2 (MJD 51300, binary phase 0.77), we obtained $K$-band
(2.2 $\mu$m) spectroscopic observations using the MPE near-infrared
imaging spectrograph 3D \cite{wkk+96} on the 3.9-m Anglo-Australian
Telescope (AAT), pointing at the position determined from the ATCA
observation.  This observation revealed a bright star with $K$-band
magnitude of 10.05$\pm$0.05 within 1\arcsec\, of the nominal position.
The observations were made using the tip-tilt correction system {\sc
rogue} \cite{tkw+95}.  3D uses an integral field unit to split the
light from a spatial grid of $16\times16$ 0.4 arcsec pixels into 256
separate spectra.  The effective resolution, $\lambda/\Delta\lambda$,
of 1100 is obtained by combining two spectra whose wavelength centre
is shifted by half a pixel using a piezo-driven flat mirror.  We
observed HD161840 (spectral type B8V) as an atmospheric standard.  A
Lorentzian was fitted to the hydrogen Brackett-$\gamma$ absorption
during reduction.  We also observed HD169101 (spectral type A0V) as a
flux calibrator, assuming $V-K=0$.  Data were reduced using routines
written for the {\tt GIPSY} data reduction package \cite{vtb+92}. The
reduction sequence involved flat fielding, bad pixel correction,
merging of the separate images corresponding to the two sub-spectra,
wavelength calibration, and formation of a data cube.  The final
spectrum of the source, which had an effective integration time per
pixel of 1680 seconds, was then extracted from the merged cube, and
divided by the atmospheric standard.

At the same time, we obtained a $2^{\prime}\times2^{\prime}$ $K$-band
image of the field using the near-infrared array camera CASPIR
\cite{mhd+94} on the Australian National University 2.3-m telescope 
at Siding Spring Observatory.  The exposure time was 1 minute, and the
limiting K magnitude was $\sim17$.  This image is shown in
Fig.~\ref{fig:caspir}.

The bright star from the 3D observation was identified on the CASPIR
image as shown in Fig.~\ref{fig:caspir}, with a consistent $K$-band
magnitude of 10.03$\pm$0.01.  Astrometric reduction of this image was
performed at the Royal Observatory Edinburgh. Twelve secondary
reference stars were identified on both the $K$-band image and an
archival UK Schmidt Telescope $R$-band plate. Positions of these stars
and hence the candidate star were related to the Hipparcos reference
frame using the Tycho-AC catalogue \cite{ucw98} and SuperCOSMOS
digitised images \cite{hmm+98}. This yielded a position for the star
of R.A.(2000) $17^{\rm h}40^{\rm m}50\fs01(1)$, Dec.(2000)
$-30^{\circ}52\arcmin 03\farcs 8(2)$. Within the combined errors, this
position agrees with the position of the pulsar given in
Table~\ref{tab:parms}.  A simple count yields a density of objects
bright enough to be detected in the CASPIR image of roughly $0.015$
per arcsec$^2$.  Using an error region of 0.86~arcsec$^2$, which
includes the 95\% confidence regions for both the radio and optical
positions, the probability of a chance coincidence is 1.3\%.

We note that the candidate companion star is included in the Point
Source Catalogue of the Two Micron All Sky Survey (2MASS)
collaboration, with the designation 2MASSI~1740500$-$305204.  In an
observation on 1998 Aug. 14 (binary phase 0.64), the survey determined
$J$, $H$ and $K$ apparent magnitudes of 14.523$\pm$0.046,
11.441$\pm$0.024, and 10.009$\pm$0.030, respectively.  Within the
uncertainties, therefore, the star appears to be stable in magnitude.

The $K$-band spectrum obtained with the AAT is shown in
Fig.~\ref{fig:spec}, with the most prominent lines indicated.  There
is significant absorption of metals and both $^{12}$CO and $^{13}$CO.
We have compared the spectrum with those of catalogue sources
\cite{kh86,wh97a} by eye and by calculating equivalent widths.
The results indicate that the star is likely to have a spectral type
between K5 and M3.

Figure~\ref{fig:spec} also indicates hydrogen Brackett-$\gamma$ in
emission. The presence of this line in late-type stars is usually
taken to indicate the presence of a compact companion with a hot
accretion disk providing the ionising flux, for example, as in the
X-ray binary GX1+4 (Chakrabarty \& Roche 1997; Davidsen, Malina \&
Bowyer 1977)\nocite{cr97,dmb77}. In the case of PSR~J1740$-$3052, we
believe there is no accretion disk (see \S\ref{sec:peri_xray} below),
and so the heating photons must have another source. In principle,
they could come from a shock at the interface between the pulsar and
supergiant winds.  However, at the nominal distance of 11\,kpc, the
observed strength of the Brackett-$\gamma$ is 20\% of the pulsar's
spin-down luminosity; as Brackett-$\gamma$ is only one of many
recombination lines of ionised hydrogen, we argue that the pulsar
cannot provide sufficient energy to produce the observed line
strength.

Because of this dilemma, further $K$-band spectroscopic observations
of the late-type star were made using CASPIR on the 2.3-m telescope at
Siding Spring Observatory on 2000 Nov. 4 (binary phase 0.16). The A0
star BS6575 was used as flux calibrator, with interpolation over
Brackett-$\gamma$ during reduction. This observation gave a spectrum
similar in appearance to that in Fig.~\ref{fig:spec} but with no
significant indication of Brackett-$\gamma$. As a further check, a K5
star (BS6842) and an M1 star (BS6587) were used as calibration for the
supergiant spectrum -- in no case was significant Brackett-$\gamma$
seen.  It therefore appears that the Brackett-$\gamma$ emission seen
in Fig.~\ref{fig:spec} is highly variable.  Variable Brackett-$\gamma$
emission is seen in late-type stars which vary in magnitude
{\cite{lw00} but it is difficult to explain in an apparently stable
star such as this one.  Perhaps the star is indeed variable, and the
three different observations happen to have been taken at the same
pulsational phase.  We note that the spacing in days between the two
CASPIR observations is roughly twice that between the 2MASS
observation and the initial CASPIR/3D observation.

The 2MASS observations can be used to determine the extinction toward
the star and hence its bolometric magnitude.  For spectral types
ranging from K5 to M3, the intrinsic $J-K$ colour should be between
0.99 and 1.12, and the intrinsic $H-K$ colour between 0.19 and 0.25
\cite{hbsw00}.  Using the standard universal extinction law
\cite{rl85} to derive $A_K/E(J-K) = 0.66$ and $A_K/E(H-K) = 1.78$, we
find that the $K$-band extinction $A_K$ must be in the range 2.1 to
2.3 magnitudes.  The $K$-band bolometric correction is approximately
2.6 for a K5 star and 2.7 for an M3 star \cite{hbsw00}.  Thus if the
star were at the estimated pulsar distance of $\sim11$\,kpc, the
observed $K$ magnitude of about 10.03 would imply a bolometric
magnitude in the range $-4.6$ to $-4.9$.  This luminosity implies a
stellar mass of only 6 to 7$\,M_{\odot}$ for a supergiant
\cite{mm89}, not large enough to make this star the pulsar companion.

It is well recognised that distances estimated from the Taylor \&
Cordes (1993) dispersion measure model may be significantly in error.
The models of Maeder \& Meynet (1989) predict a bolometric magnitude
of $-6.5$ for a 12$\,M_{\odot}$ star, requiring the system to be at a
distance of 23\,kpc --- more than twice as far as predicted by the
dispersion measure model.  We do not believe this to be likely, but it
is not impossible.

A further problem which arises with identifying this star as the
companion, at either distance, is the stellar radius.  As noted in
\S\ref{sec:soln}, red supergiants have radii of several
hundred $R_{\odot}$ \cite{vlt+99}, larger than permitted by the pulsar
orbit.  In fact, requiring that the companion be contained within the
Roche lobe forces the stellar radius to be less than 0.4\,AU.  In
short, the luminosity and radius required of a late-type supergiant
make it quite unlikely that this star is the pulsar companion.  If the
observed star lies near the Galactic centre at a distance of 8.5\,kpc,
its bolometric luminosity would be approximately $-4.3$, consistent
with that of the roughly solar-mass red giants which dominate the
Galactic Bulge stellar population.  We believe that this is the most
self-consistent explanation of the properties of this star.

It therefore appears that the positional agreement between the
late-type star and the pulsar is indeed a coincidence.  If there were
to be a B~star hidden by the light of the late-type star and at the
estimated pulsar distance of 11\,kpc, its $K$-band magnitude would be
roughly $15$ and it would not appreciably change the observed $K$-band
spectrum.

\section{Observations near Periastron}\label{sec:peri}

In a further attempt to distinguish between a black hole and a
non-degenerate companion, we undertook dual-frequency radio monitoring
campaigns of the pulsar around the periastrons of 2000 February 10
(MJD 51584.5) and 2000 September 28 (MJD 51815.5).  Observations were
made at Parkes on most days in the periods 2000 February 5 -- 17 and
2000 September 19 -- 30 at centre frequencies of 660\,MHz and
1390\,MHz.  The goals of these campaigns were to verify the absence of
eclipses and to look for evidence of variations in the pulsar's
dispersion and rotation measures, as such changes might be expected
from the interaction of the pulsar signal with the wind from a
non-degenerate companion.

\subsection{Timing Observations}\label{sec:peri_time}

The Parkes dual-frequency timing observations were performed at
660\,MHz using a $2\times256\times0.125$-MHz filterbank and, using the
central beam of the multibeam receiver, at 1390\,MHz with a
$2\times512\times0.5$-MHz filterbank, both employing the 1-bit
digitisation system described in \S\ref{sec:obs}. The pulsar was
detected on each observing day, demonstrating conclusively that there
is no eclipse.  Fig.~\ref{fig:dmvar} shows timing residuals of the
660-MHz and 1390-MHz data as a function of orbital phase, with the
dispersion measure held constant at the value given in
Table~\ref{tab:parms}.  There is clear evidence for increased and
variable dispersion before periastron.  This pattern is consistent
with the observed longitude of periastron, as the companion is between
us and the pulsar before periastron and beyond the pulsar after
periastron.  Therefore, one would expect increased DMs before
periastron and more stable values after.

These observations show that the increase in DM on any
given day relative to the reference DM of 740.9(2)\,cm$^{-3}$\,pc is
typically of the order of 1 or 2\,cm$^{-3}$\,pc.  For the observations
before periastron, the extra distance travelled across the pulsar orbit
is about $10^{13}$ cm, or 0.67 AU, leading to an estimated electron
density inside the orbit of a few $\times 10^{5}$\,cm$^{-3}$.

We consider once again the possibility of a late-type, cool, bright
star as the companion.  The expected stellar wind from such a star is
of the order of $10^{-6}\,M_{\odot}\,{\rm yr}^{-1}$ (e.g., Dupree
1986) \nocite{dup86} with an ionised fraction of 0.002 -- 0.02
\cite{dl83}.  The closest approach of the two stars for $i = 70\degr$,
$m_2 = 12.5\,M_{\odot}$ is only 0.75\,AU, just twice the maximum
possible stellar radius based on Roche-lobe considerations.  Assuming
a distance from the companion centre of about 0.75 AU and a stellar
wind velocity of $\sim 30$\,km\,s$^{-1}$ \cite{dup86}, we arrive at an
estimate for an ionised mass-loss rate of a few $\times
10^{-11}\,M_{\odot}{\rm yr}^{-1}$ and therefore an overall mass-loss
rate of $10^{-9}$--$10^{-8}\,M_{\odot}{\rm yr}^{-1}$.  This is smaller
by two orders of magnitude than the expected rate for these stars,
which could possibly be explained if the mass loss is very clumpy as
found, for example, in $\alpha$ Ori (e.g., Skinner \& Whitmore
1987). \nocite{sw87a} However, we believe that this low mass-loss rate
casts further doubt on the association of the late-type star with the
pulsar.

In the case of an early B~star, we follow the arguments of Kaspi et
al. (1996) \nocite{ktm96} and references therein in adopting the
following law for the wind velocity $v_w$:
\begin{equation}
v_w(r) = v_{\infty}(1-R_2/r)^{1/2},
\end{equation}
where $r$ is the distance from the centre of mass of the star and
$v_{\infty}$ is 1--3 times the escape velocity $v_{\rm esc} \approx
725\,$km\,s$^{-1}$.  The electron density $n_e(r)$ at any point $r$
from the star can be found from mass conservation:
\begin{equation}
\dot M = 4\pi r^2 n_e(r) m_p v_w(r)
\end{equation}
where $\dot M$ is the mass-loss rate and $m_p$ is the proton mass.
For the observations just before periastron, we integrate numerically
along the line of sight through the pulsar orbit in the following manner:
\begin{equation}
I = \int \frac{1}{\sqrt{1-R_2/r}}\frac{1}{r^2}\,dl
\end{equation}
 in order to find the expected difference in DM.  The end result is:
\begin{equation}
\dot M = 1.1\times 10^{-9} \left(\frac{v_{\infty}}{v_{\rm esc}}\right) 
\frac{\Delta\,{\rm DM}}{I} M_{\odot}\,{\rm yr}^{-1}
\end{equation}
where $\Delta$DM is the difference in DM before and after periastron
in units of cm$^{-3}$\,pc and $I$ is in units of AU$^{-1}$.  Both
$\Delta$DM and $I$ are of the order of unity here, implying a
mass-loss rate of a few $\times 10^{-9}\,M_{\odot}{\rm yr}^{-1}$,
roughly what is predicted for early-type stars of this mass in the
Galaxy (e.g., de Jager et al. 1988). \nocite{dnv88} We note that this
mass-loss rate is two orders of magnitude higher than the upper limit
found for PSR~J0045$-$7319 in the Small Magellanic Cloud, lending
support to the argument for a metallicity dependence of the mass-loss
rate for such stars \cite{ktm96}.  

We conclude that the observed DM variations are better explained by an
early B-star companion rather than by a late-type supergiant.

\subsection{Polarimetric Observations}\label{sec:peri_pol}

The polarisation of the mean pulse profile was measured on a total of
18 days before, during and after the 2000 February periastron, and on
several other occasions throughout the orbit, using the centre beam of
the multibeam receiver and the Caltech correlator
\cite{nav94}. The centre frequency and bandwidth were 1318.5 MHz and
128 MHz, respectively. Observations were made in pairs at orthogonal
position angles, typically for 30 min at each angle; summing of these
orthogonal pairs removes most of the effects of instrumental
polarisation. Data were calibrated following standard procedures
\cite{nms+97}, except that the full frequency resolution of the
correlator, $128 \times 1$ MHz, was retained during the processing.

In general, the mean pulse profile is weakly polarised.
Fig.~\ref{fig:poln} shows the mean profile resulting from adding all
of the data obtained over the periastron period, between 2000 February
4 and February 17, a total of 5.85 hours of observation. The average
linear polarisation $\langle L\rangle/\langle I\rangle$, where $L =
(Q^2 + U^2)^{1/2}$, is only $1 \pm 4$\%. However, there is some
significant circular polarisation, with a hint of a sense reversal
near the pulse peak. This suggests that the very narrow pulse (full
width at half-maximum of 8\degr~ of longitude) is from the core region
of the polar cap (e.g., Rankin 1983). \nocite{ran83}

It is possible that the very low linear polarisation is due to Faraday
depolarisation in the wind of the companion. This effect is seen in
the eclipsing Be-star system PSR B1259$-$63 \cite{jml+96}. To
investigate this we summed the individual channel data for each
orthogonal pair over a range of rotation measures (RM) from $-$5000 to
+5000 rad m$^{-2}$ in steps of 50 rad m$^{-2}$.  Where significant
linear polarisation was found, an improved value of the rotation
measure was computed from a weighted mean position angle difference
between the two halves of the observed band.  On three of 20 or so
observations made away from periastron, on 2000 February 28, March 26
and May 31, significant linear polarisation (20 -- 30\%) was observed
with rotation measures of +180, $-$85, and $-$220 rad m$^{-2}$,
respectively.  No significant polarisation ($< 10$\%) was observed at
any rotation measure within the search range during the periastron
period or on other occasions.

These results suggest that Faraday rotation is occurring in the wind
of the companion star and that it is highly variable.  Combined with
the DM variations of 1 or 2 cm$^{-3}$ pc during periastron passage
(\S\ref{sec:peri_time}), the RM changes indicate that the magnetic field
strength in the wind region (weighted by the local electron density)
is at least a few times 0.1 mG. The observed variations suggest that
the field structure is complex, and so this value, which is integrated
along the line of sight, is a lower limit to the actual field strength
in the wind region.

The very existence of DM and RM variations appears to argue against a
black hole as the pulsar companion, although it is perhaps plausible
that the passage of the signal through the extended atmosphere of the
late-type star could be responsible for the variations if the late-type 
star is foreground and the geometry is favourable.

\subsection{Archival X-ray Observations}\label{sec:peri_xray}

For a non-degenerate companion, particularly an extended giant star,
accretion of companion wind material onto the neutron star could in
principle occur near periastron, where the distance of closest
approach is 0.72\,AU/$\sin i$.  Such accretion might result in
observable X-ray emission.

To investigate this possibility, we examined archival X-ray
observations of the field near PSR~J1740$-$3052.  Serendipitously, the
{\it ASCA} X-ray telescope \cite{tih94} observed a field containing
PSR~J1740$-$3052 on 1995 September 26 (Sequence ID 53016050) as part
of its survey of Galactic Ridge emission.  This is only 19~days after
a periastron passage.  In addition, the source was only 8$'$ from the
centre of the field of view in the observation.

We have reduced data from the two co-aligned Gas Imaging Spectrometers
(GIS) onboard {\it ASCA}.  The effective total GIS exposure was $2
\times 12.5$~ks.  We used the standard {\it ASCA} data analysis tool
{\tt XSELECT} to produce a first image of the field, for both GIS2 and
GIS3.  For this image, we included counts having energies between 2
and 10\,keV, as softer emission is likely to have been absorbed, given
the large expected column density toward the source (see below).  The
resulting image was exposure-corrected, and was further corrected for
variations in the particle background over the fields-of-view using
the {\tt FTOOL} {\tt ascaexpo}.  For details, see Roberts, Romani \&
Kawai (2001).\nocite{rrk01} The corrected GIS2 and GIS3 images were
combined and smoothed with a Gaussian function having FWHM
$50\arcsec$.

No significant emission was detected from the pulsar position.  To set
an upper limit on the flux, we first found the rms scatter of the 
image, $1.5 \times 10^{-7}$~counts~s$^{-1}$\,cm$^{-2}$~pixel$^{-1}$.
This number was then multiplied by the number of pixels (117) in the
half-power region of the {\it ASCA} point-spread function.
Multiplying by 2 for the full power, and by 3 to yield a $3\sigma$ upper
limit, we find that the observed flux from the source in the 2--10~keV
band is $< 1 \times 10^{-4}$~counts~s$^{-1}$\,cm$^{-2}$.

In order to determine the upper limit on the energy flux and hence
source luminosity, we assume a simple power-law model having photon
index $-2$, and an equivalent neutral-hydrogen absorbing column of
$N_H \simeq 3 \times 10^{22}$\,cm$^{-2}$.  
We use {\tt HEASARC}'s tools\footnote{{\tt WebSpec}:
http://heasarc.gsfc.nasa.gov/cgi-bin/webspec and {\tt W3PIMMS}:
http://heasarc.gsfc.nasa.gov/Tools/w3pimms.html.} to convert the above
upper limit on the photon flux into an upper limit on the unabsorbed
energy flux, $1.4 \times 10^{-12}$~erg~s$^{-1}$~cm$^{-2}$, in the
2--10~keV band.  For a distance of 11\,kpc as estimated from the
pulsar's dispersion measure (see \S2.3), this implies an upper limit
on the source's 2--10~keV luminosity $L_x < 2 \times
10^{34}$~erg~s$^{-1}$.  Assuming the simplest possible accretion
model, this implies $\dot{M} < L_x R_p / G m_1 = 1.8 \times 10^{-12}
\; M_{\odot}$~yr$^{-1}$, where $R_p = 10$~km and $m_1 = 1.35 \,
M_{\odot}$ are the assumed neutron star radius and mass, respectively.

Given this upper limit, accretion is very unlikely to have occurred.
Such a low $\dot{M}$ is unlikely to have sufficient pressure to
overcome the pulsar wind pressure (see, e.g., Tavani, Arons \& Kaspi
1994), \nocite{tak94} so material is unlikely to have come within the
accretion radius.  We note however that our upper limit does not
preclude the existence of non-thermal shock-powered X-rays like those
seen in the pulsar/Be star binary PSR~B1259$-$63 near periastron
\cite{ktn+95,hnt+96}.  The shock emission is ultimately powered by the
pulsar's spin down.  PSR~J1740$-$3052 has a much smaller spin-down
luminosity than PSR~B1259$-$63 and is much more distant.  Even in the
unlikely event that {\it all} of the spin-down luminosity of
PSR~J1740$-$3052 ($\dot{E} = 5.5 \times 10^{33}$~erg~s$^{-1}$, see
Table~\ref{tab:parms}) were converted into shock emission in the X-ray
band, it would be unobservable in the archival {\it ASCA} data.

\section{Discussion}\label{sec:disc}

On evolutionary grounds, there is no reason to prefer one type of
candidate companion over another.  As we have discussed above,
PSR~J1740$-$3052 is a young pulsar which has not undergone an episode
of mass transfer from its companion.  The pulsar's characteristic age
is $3.5 \times 10^5$\,yr, while a late-type supergiant star might be
expected to evolve from an OB star in $\sim10^{7}$\,yr, making either
type of non-degenerate companion consistent from the point of view of
stellar ages.  The fact that the system remained bound on formation of
the neutron star suggests that the pre-supernova star was the less
massive of the two at the time of the explosion. This is consistent
with evolutionary scenarios involving mass transfer on to the
initially lighter star \cite{py98}.  This mass transfer may also have
somewhat accelerated the companion's evolution.  Alternatively, a
suitably oriented kick on formation could have kept the system bound
\cite{bai88,tt98}.  In the case of a black-hole companion, the neutron
star would be the second-formed compact object, as the black-hole
progenitor would have been more massive initially.

Two pulsar systems are currently known to have non-degenerate
companion stars, PSRs~B1259$-$63 \cite{jml+92} and J0045$-$7319
\cite{kjb+94}, and these contain Be and B stars, respectively.  These
systems are considered likely progenitors of HMXBs.  If, as we believe
is likely, the companion to PSR~J1740$-$3052 is also an early B star,
then this system will likely also become an HMXB in the future, as the
neutron star begins to accrete matter from the evolving companion's
wind.  As the companion evolves to overflow its Roche lobe, the system
will enter a common-envelope phase and the neutron star will begin to
spiral in.  The current orbital period of this system, 231 days, makes
the outcome after this point uncertain \cite{vdh93}.  The neutron star
may spiral in completely, resulting in a red giant star with a neutron
star core: a Thorne-Zytkow object \cite{tz77}.  Alternatively, there
may be enough energy released during the orbital spiral-in to eject
the common envelope, leaving behind the evolved core of the companion.
Given the current mass of the companion, this core is likely to
undergo a supernova explosion itself, leaving behind either a bound
double-neutron-star system such as PSR~B1913+16, or else two isolated
neutron stars, one mildly recycled, one young.

We have argued throughout this paper that the companion is most likely
to be an early-type B star rather than a black hole or the late-type
star which is coincident with the pulsar's position.  Our arguments
may be summarized as follows:
\begin{enumerate}

\item The magnitude of the late-type star coincident with 
the pulsar position can be made consistent with the bolometric
magnitude of an 11\,$M_{\odot}$ star only if the dispersion measure
estimate of the distance to the pulsar is low by a factor of two.  In
contrast, the colours and magnitude of the star are perfectly
consistent with an AGB star of about 1\,$M_{\odot}$ at the Galactic
centre.

\item An early-B main sequence star at the nominal pulsar distance of
about 11\,kpc and hidden by the late-type star would not significantly
alter the observed $K$-band magnitude or spectrum.

\item The radii of late-type supergiant stars are as large as or larger 
than the pulsar's orbit.  No significant mass has been transferred to
the pulsar from the companion, requiring the companion to be confined
inside its Roche lobe of radius roughly 0.4\,AU, improbably small for
a late-type supergiant.  Such a small radius would also require either
a higher temperature or a smaller luminosity for the star, contrary to
our understanding of the evolution of these objects \cite{mm89}.

\item Even for a late-type supergiant of the small required radius, the
calculated magnitudes of the tidal and spin quadrupoles predict an
advance of periastron in the pulsar's orbit an order of magnitude
larger than that observed.  By contrast, similar estimates for an
early B~star predict values not much larger than the
general-relativistic prediction, and a good match to the observations.
A black-hole companion is also consistent with these observations.

\item The existence of orbital-phase-dependent DM and RM variations 
argues for a non-degenerate companion of some kind, and against a
black-hole companion unless there is a fortuitous alignment between
the pulsar's orbit and the extended wind of the foreground late-type
star.  Furthermore, the observed magnitude of the DM variations
implies a stellar wind two orders of magnitude smaller than that
predicted for late-type stars (e.g., Dupree 1986) but consistent with
that expected for an early B~star.
\end{enumerate}

All points considered, we find that the bulk of the evidence points to
a non-degenerate companion, but to an early B~star rather than to the
late-type star observed to be coincident with the pulsar position.  It
should be possible to establish whether or not the late-type star is
the companion through a careful search for Doppler radial velocity
variations in the stellar spectrum.  The expected total range of
radial velocity variation is 22\,km\,s$^{-1}$ for an $11\,M_{\odot}$
companion, or 15\,km\,s$^{-1}$ for a $16\,M_{\odot}$ companion; this
should be measurable with a high-resolution spectrometer.  Further
multifrequency radio monitoring of the orbital DM and RM variations
will lead to a characterization of the wind of a non-degenerate
companion.  Finally, continued long-term timing of the pulsar will
lead to precise values of $\dot\omega$ and $\dot x$, allowing a
separation of general-relativistic effects from those caused by the
quadrupole of a non-degenerate companion, thus providing final proof
of the nature of the companion.

\section*{ACKNOWLEDGEMENTS}
We thank Steve Eikenberry, Anita Richards and Lars Bildsten for
helpful discussions, Mallory Roberts for help with the X-ray
reduction, and George Hobbs for help with observing.  The Parkes radio
telescope is part of the Australia Telescope which is funded by the
Commonwealth of Australia for operation as a National Facility managed
by CSIRO. IHS received support from NSERC and Jansky postdoctoral
fellowships.  VMK is an Alfred P. Sloan Research Fellow, and received
support from an NSF CAREER award (AST-9875897), NASA grant NAG5-9120
and an NSERC grant (RGPIN228738-00).  F. Camilo is supported by NASA
grant NAG5-9095.  This research has made use of the Astronomical Data
Center (ADC) at NASA Goddard Space Flight Center, and the Simbad and
Vizier services operated by CDS Strasbourg.  It has also used data
products from the Two Micron All Sky Survey, which is a joint project
of the University of Massachusetts and the Infrared Processing and
Analysis Center/California Institute of Technology, funded by the
National Aeronautics and Space Administration and the National Science
Foundation.  The Digitised Sky Surveys were produced at the Space
Telescope Science Institute under U.S. Government grant NAG W-2166.

\pagebreak

\begin{table}
\begin{center} 
\caption{Parameters of PSR~J1740$-$3052 \label{tab:parms}}
\begin{tabular}{lc} 
\hline
\hline
\multicolumn{2}{c}{Measured Parameters}     \\                    

Right Ascension (J2000)$^a$ & $17^{\rm h}40^{\rm m}50\fs031(5)$ \\
Declination (J2000)$^a$ & $-30\degr52\arcmin 04\farcs1(3)$\\ 
Dispersion Measure (cm$^{-3}$ pc) & 740.9(2) \\ 
Period (s) & 0.570309580513(16) \\
Period Derivative & $2.54969(4)\times 10^{-14}$ \\ 
Epoch of Period (MJD) & 51452.0 \\ 
Orbital Period (days) & 231.02965(3) \\ 
Projected Semi-major Axis (light-seconds) & 756.9087(4) \\ 
Eccentricity & 0.5788720(4) \\
Epoch of Periastron (MJD) & 51353.51233(3) \\ 
Longitude of Periastron (degrees) & 178.64613(6) \\ 
Advance of Periastron (degrees\,yr$^{-1}$) & 0.00021(7) \\ 
Derivative of Projected Semi-major Axis$^b$ & $<4 \times 10^{-11}$  \\
Derivative of Orbital Period$^b$ & $<4 \times 10^{-8}$  \\
Data Span (MJD) & 51032 -- 51872 \\
R.M.S. Timing Residual (ms) & 0.8 \\ 
Flux Density at 1400 MHz (mJy) & 0.7(2) \\ \\ 
\hline
\multicolumn{2}{c}{Derived Parameters} \\
Galactic Longitude (degrees)            & 357.8 \\
Galactic Latitude (degrees)             & $-$0.13 \\
Distance (kpc)                          & $\sim 11$ \\
Characteristic Age (yr)                 & $3.5\times 10^5$       \\
Surface Magnetic Field (G)              & $3.9\times 10^{12}$    \\
Rate of Energy Loss (erg\,s$^{-1}$)     &  $5.5\times 10^{33}$ \\
Mass Function (M$_{\odot}$)             & 8.723248(12)  \\
     
\hline
\end{tabular} 
\end{center} 
$^a$Position determined from interferometric observation. \\
$^b$Fit while holding all other parameters constant at the values shown.
\end{table} 

\begin{figure}
\centerline{\psfig{file=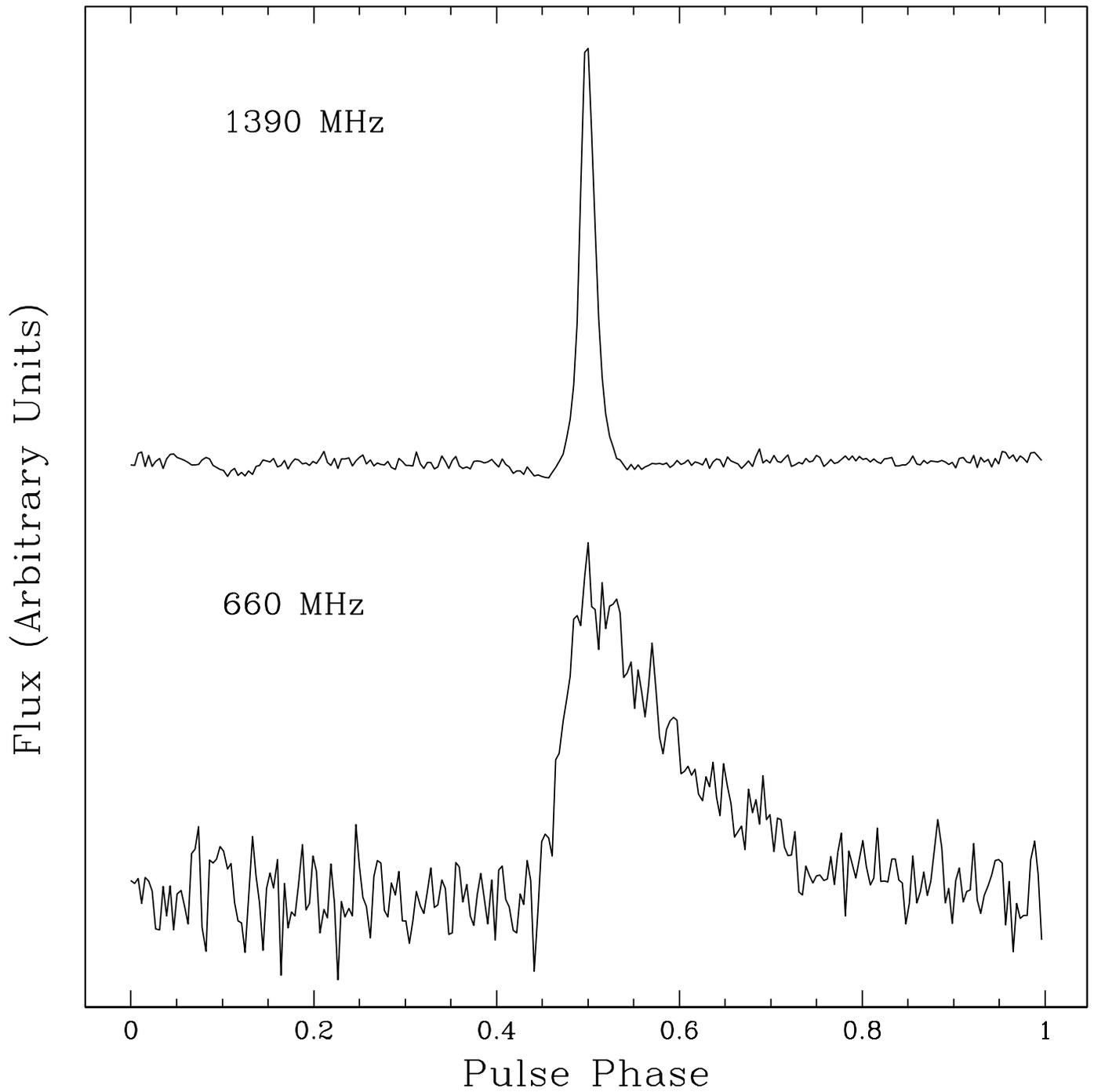}}
\caption{Mean pulse profiles at 1390\,MHz and 660\,MHz.  Integration
times were 11.2 hours at 1390\,MHz, and 11.6 hours at 660\,MHz.  The
dispersion smearing is 1.11\,ms at 1390\,MHz, and 5.36\,ms at
660\,MHz.  The small dip at the leading edge of the 1390-MHz profile
is an instrumental artefact.  The scattering timescale derived from
the 660-MHz profile is roughly 58\,ms.  This scales by $\nu^{-4.4}$
to about 9\,ms at 1\,GHz, only 25\% of the value predicted by the
Taylor \& Cordes (1993) model.}
\label{fig:profs} 
\end{figure}

\begin{figure}
\centerline{\psfig{file=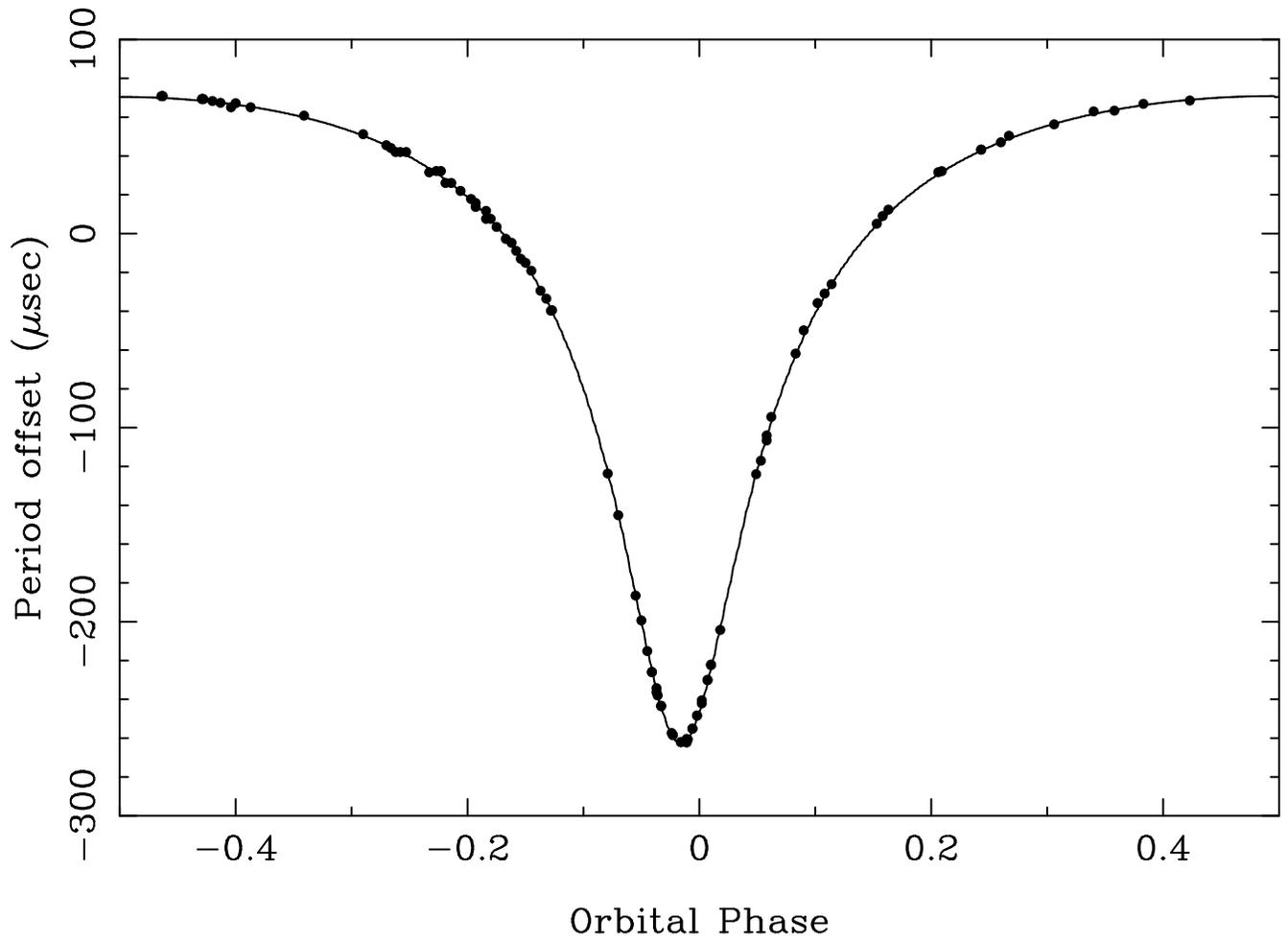}}
\caption{Observed variations of the solar-system barycentric 
period of PSR~J1740$-$3052 over the 231-day orbital period measured
using the 76-m Lovell telescope at Jodrell Bank Observatory and the
64-m telescope at Parkes. The curved line represents the fit of a
binary model to the data.  Orbital phase 0 is periastron, which
occurs nearly in the plane of the sky.}
\label{fig:perphase} 
\end{figure}

\begin{figure}
\centerline{\psfig{file=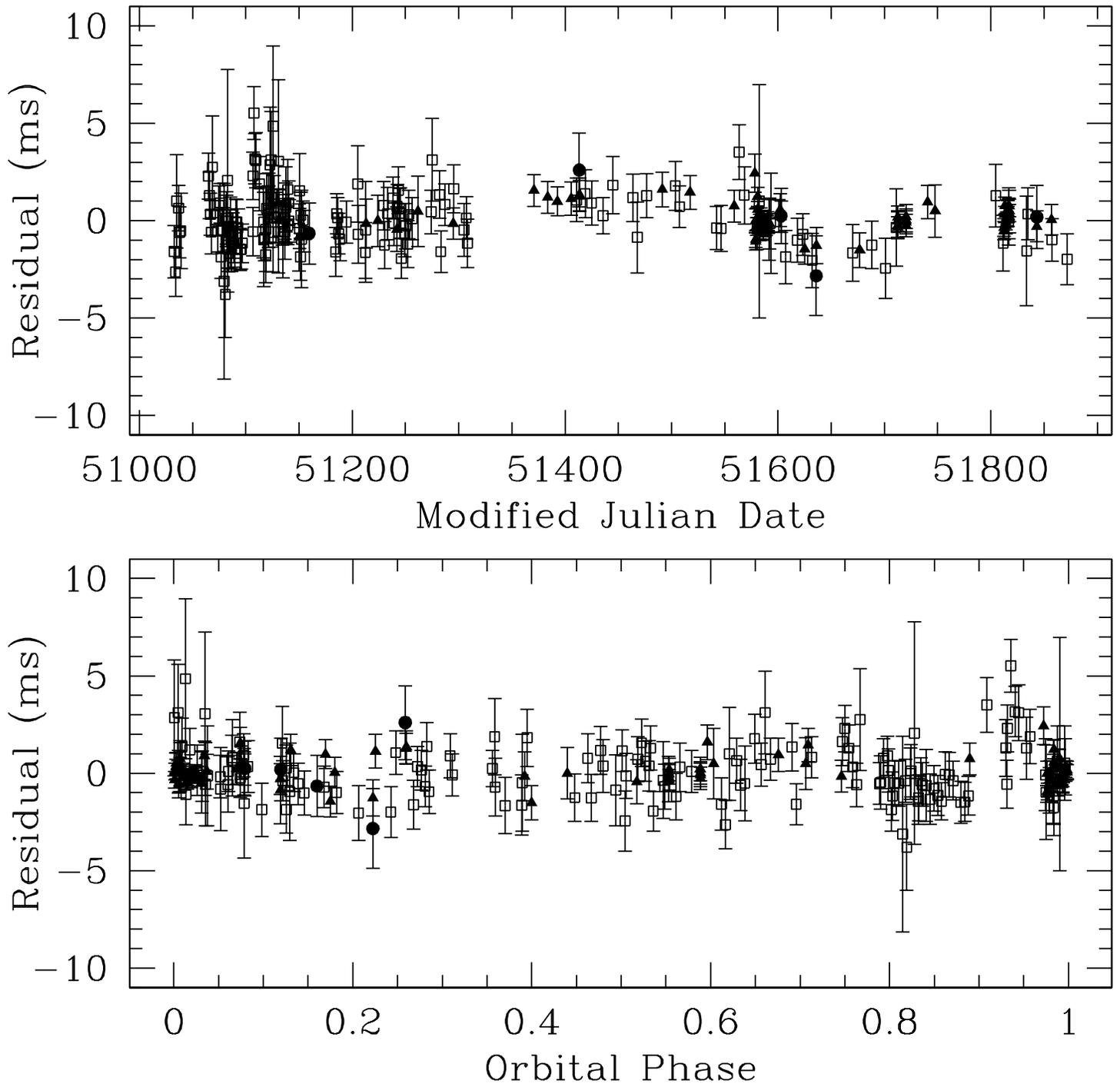}}
\caption{Timing residuals relative to the best-fit solution for 
PSR~J1740$-$3052 over the period 1998 Aug. 7 -- 2000 Nov. 23.  Open
squares represent Jodrell Bank 1400-MHz data, filled triangles Parkes
1400-MHz data and filled circles 660-MHz data.  Top panel: residuals
as a function of time.  Bottom panel: residuals as a function of
orbital phase, where phase 0/1 is periastron.}
\label{fig:resids} 
\end{figure}

\begin{figure}
\centerline{\psfig{file=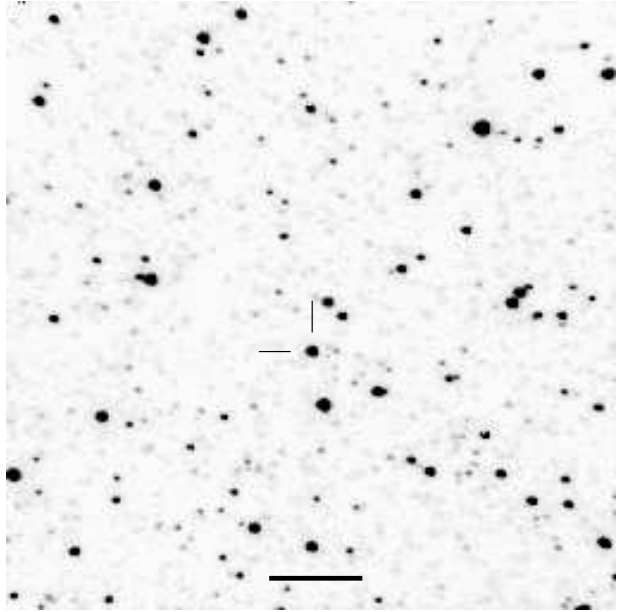}}
\caption{Image of the PSR~J1740$-$3052 field taken with
the Siding Spring Observatory 2.3-m telescope at $2.2 \mu$m
($K$-band). The coincident late-type star, near the centre of the
figure, is marked, and the thick horizontal bar is of length
$20\arcsec$. North is at the top of the figure and east to the left. }
\label{fig:caspir} 
\end{figure}

\begin{figure}
\centerline{\psfig{file=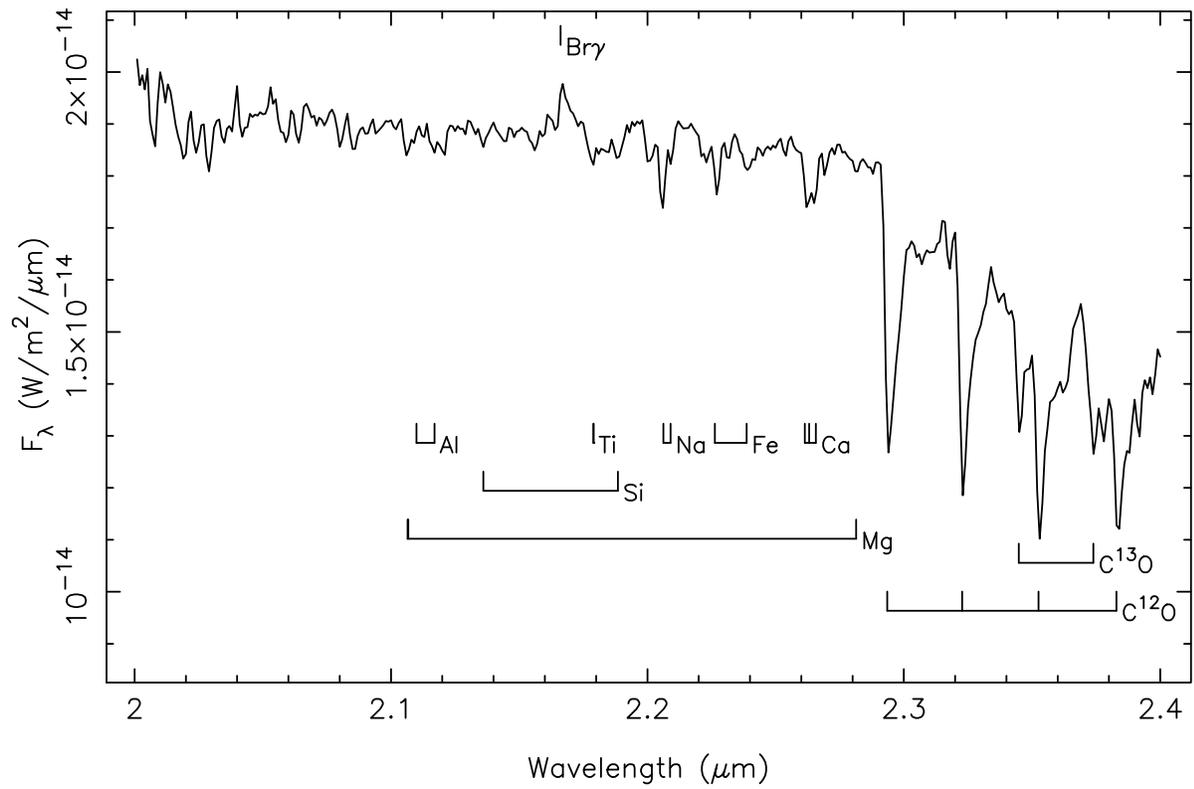,width=7.0in}}
\caption{2.2 $\mu$m ($K$-band) spectrum of the
late-type star at the position of PSR~J1740$-$3052 obtained with the
3D instrument on the Anglo-Australian Telescope.  Prominent spectral
lines are labeled.}
\label{fig:spec} 
\end{figure}

\begin{figure}
\centerline{\psfig{file=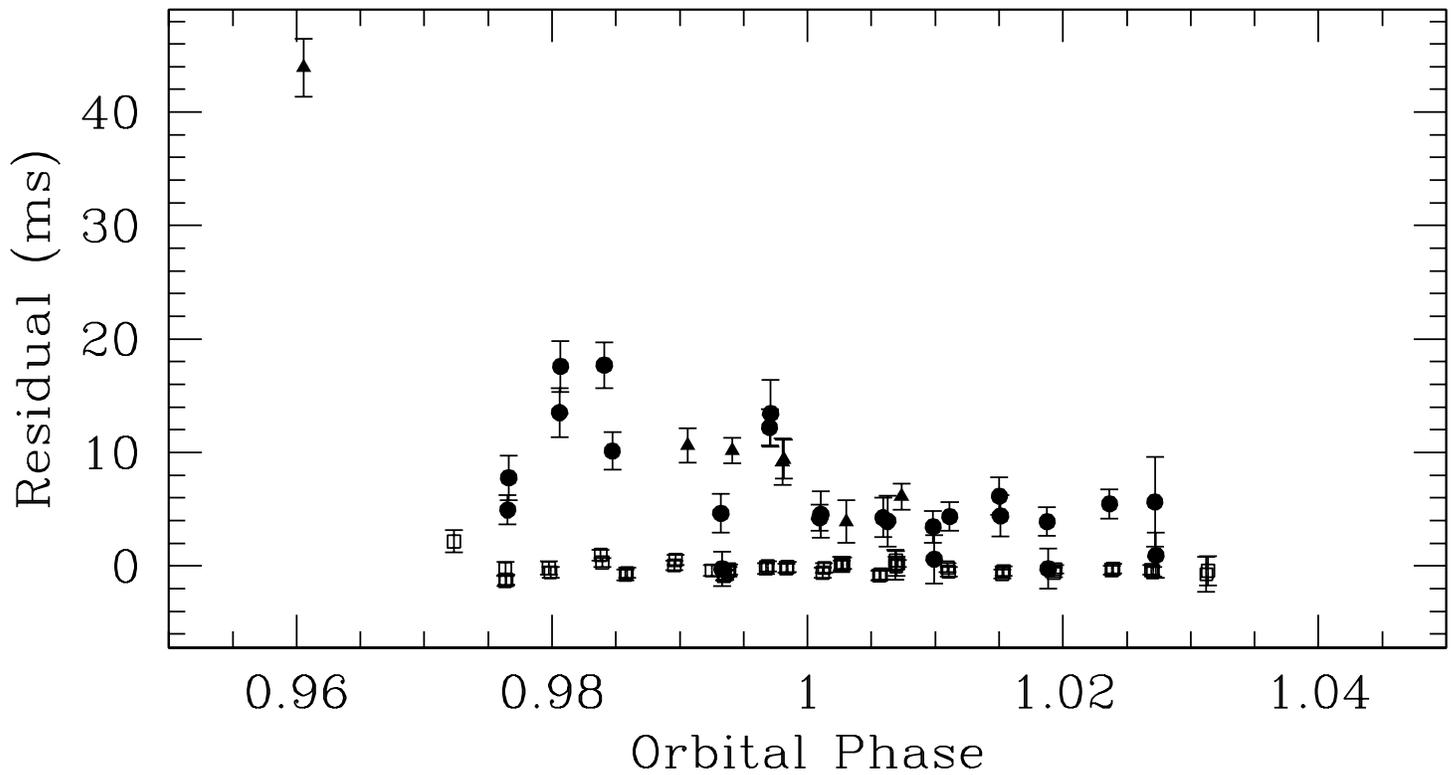}}
\caption{\label{fig:dmvar} Residuals at 660\,MHz and 1390\,MHz as a 
function of orbital phase near the periastrons of 2000 February 10 and
2000 September 28. Periastron is at orbital phase 1.0.  1390-MHz data are
indicated by open squares, 2000 February 660-MHz by filled circles, and
2000 September 660-MHz by filled triangles.  There is clear evidence for
variation in the dispersion measure before periastron.  We believe the 
660-MHz point at phase $\sim0.96$ is likely to be reliable.}
\end{figure}

\begin{figure}
\centerline{\psfig{file=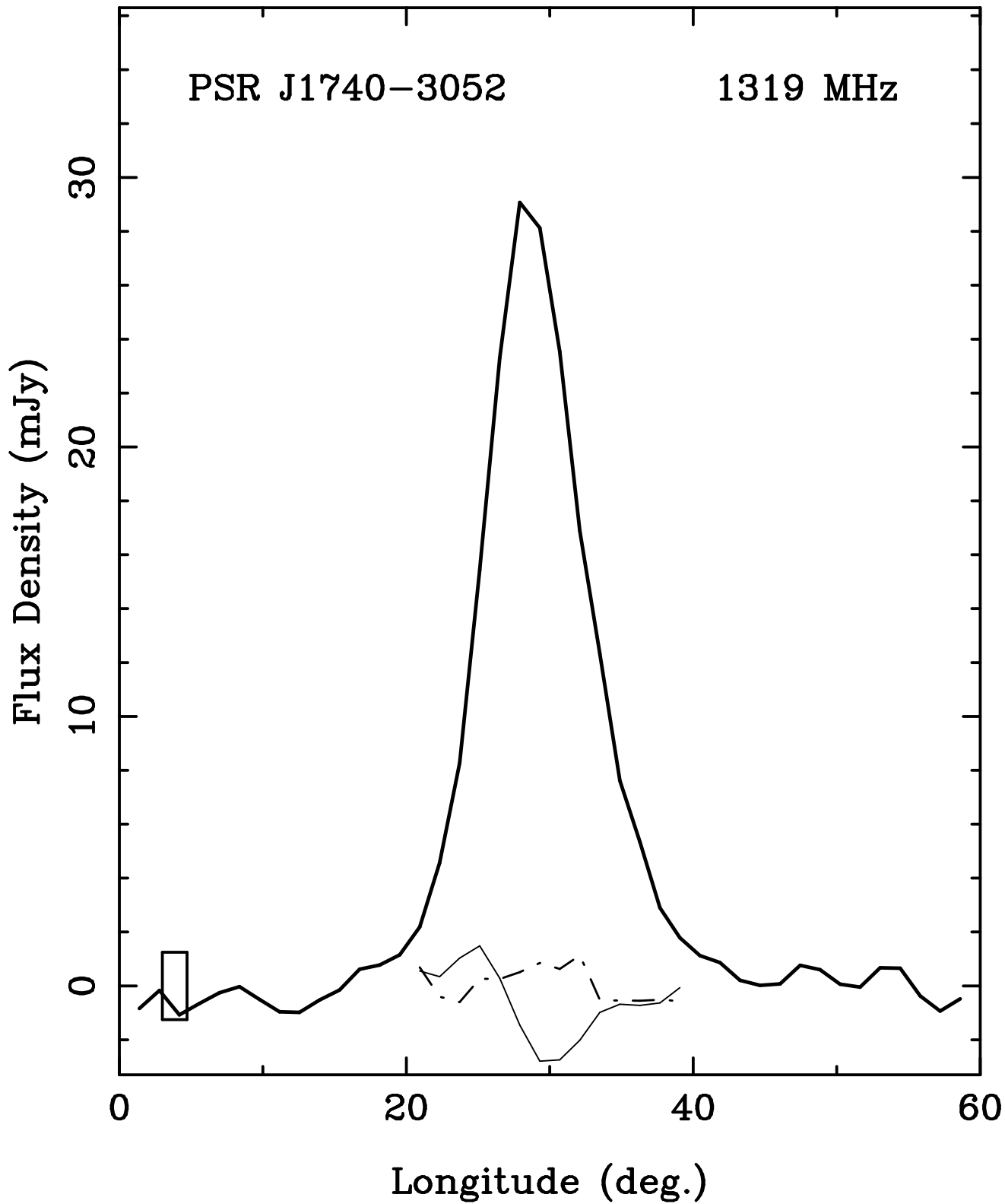}} 
\caption{Mean pulse
polarisation profile for PSR J1740$-$3052 obtained by summing 5.85 h
of data obtained at 1318.5\,MHz in 2000 February. The upper solid line
is the total intensity, Stokes $I$, the dot-dashed line is the
linearly polarised component, $L = (Q^2 + U^2)^{1/2}$, and the lower
solid line is Stokes $V = I_{LH} - I_{RH}$. The box at the left of the
baseline has a height of four times the baseline rms noise and a width
equal to the effective time resolution of the profile, including the
effects of interstellar dispersion. Position angles were not
significant, and hence are not plotted.}
\label{fig:poln} 
\end{figure}

\end{document}